\documentstyle[11pt,aaspp4,tighten]{article}

\tightenlines

\def\ch{$\chi^2$ }
\def\chs{$\chi^2$}

\begin{document}

\title{The X-Ray Spectral  Variability  of Mrk~766}

\author{Karen M. Leighly\altaffilmark{1}}
\affil{Cosmic Radiation Laboratory, RIKEN, Hirowasa 2--1, Wakoshi, 
Saitama 351, Japan}
\altaffilmark{1}{Previously an NRC fellow at NASA Goddard Space Flight Center
\author{Richard F. Mushotzky and Tahir Yaqoob\altaffilmark{2}}
\affil{Code 662.0, NASA Goddard Space Flight Center, Greenbelt, MD 20771}
\altaffilmark{2}{Also Universities Space Research Association}
\author{Hideyo Kunieda}
\affil{Department of Physics, Nagoya University, Furo-cho, Chikusa-ku,
Nagoya 464, Japan}
\author{Rick Edelson}
\affil{Department of Physics and Astronomy, University of Iowa, Iowa
City, IA 52242-1479}

\slugcomment{Submitted to the Astrophysical Journal}

% The abstract environment prints out 
% horizontal rules for the journal's editorial staff to type the
% received and accepted dates

\begin{abstract}

Analysis results from {\it ASCA} and {\it ROSAT} observations of the
narrow-line Seyfert 1 galaxy Mrk~766 are reported.  In the {\it ASCA}
observation we observed rapid variability with a doubling time scale
of 1000 seconds.  A spectral variability event was observed in which
the spectrum softened and hardened above and below $\sim 1 \rm keV$,
respectively, as the flux increased.  The spectra could be modeled
with 5 components: a power law, warm absorber, iron $K\alpha$ line and
soft excess component flux.  The spectral variability resulted from a
highly significant change in the intrinsic photon law index from
$\Gamma \sim 1.6$ to $\sim 2.0$, an increase in the warm absorber
ionization, and a marginally significant decrease in the soft
component normalization.  A $\sim 100\rm eV$ equivalent width narrow
iron $K\alpha$ line was detected in the high state spectrum.  Spectral
hardening during flux increases was observed in three {\it ROSAT}
observations.

The change in intrinsic photon index and disappearance of the soft
excess component in the {\it ASCA} spectra can be explained as a
transition from a first order pair reprocessed spectrum to a pair
cascade brought about by a sudden increase in the injected electron
Lorentz factor.  The change in the ionization of the warm absorber,
though model dependent, could correspond to the increase in flux at
the oxygen edges resulting from the spectral index change. The {\it
ROSAT} spectral variability can be interpreted by variable intensity
hard power law and a relatively nonvarying soft component, possibly
primary disk emission.  These results are compared with those reported
from other narrow-line Seyfert 1 galaxies.

\end{abstract}


\section{Introduction}

Mrk~766 is a bright ($F_{(2-10)}\sim 2 \times 10^{-11} \rm erg\,
cm^{-2} s^{-1}$), soft ($\Gamma_{0.1-2.4} \sim 2.7$) X-ray source at
redshift $z=0.012$.  The spectrum measured with the {\it Einstein} IPC
and MPC was complex and ultra-soft ($\Gamma=1.77$; $kT=18.6 \rm eV$;
Urry {\it et al.}
\markcite{34} 1990).  A shortest time scale of variability of 1000
seconds and a steep and variable power law index was found in a long
observation using {\it EXOSAT} (Molendi, Maccacaro \& Schaeidt
\markcite{20} 1993).  During the {\it ROSAT} All Sky Survey, 
Mrk~766 was bright ($F_{0.1-2.4} \sim 1.5
\times 10^{-10} \rm erg\, cm^{-2} s^{-1}$ (unabsorbed)) and
variability by a factor of three with no accompanying spectral
variability was observed in 10--12 hours (Molendi, Maccacaro \&
Schaeidt \markcite{20} 1993).  Pointed {\it ROSAT} observations
revealed spectral variability that Netzer, Turner
\& George (1994) \markcite{24} showed could not be explained by a
change in ionization of a warm absorber, and Molendi \& Maccacaro
(1994) \markcite{19} attributed to a change in the accretion rate.

Mrk~766 is a member of the X-ray narrow line Seyfert 1 (NLS1) galaxy
class (Osterbrock \& Pogge \markcite{25} 1985; Goodrich \markcite{13}
1989).  {\it ROSAT} observations of NLS1s find soft 0.1--2.4 keV X-ray
spectra and rapid, large amplitude soft X-ray variability.  The soft
X-ray spectra of NLS1s are systematically steeper than the spectra of
broad-line Seyfert 1 galaxies (Boller, Brandt \& Fink \markcite{3}
1996 and references therein). A harder high energy power law component
generally was not observed in the relatively soft {\it ROSAT} band.
Only a few observations at higher energies have been reported.  The
{\it ASCA} spectrum of the NLS1 object IRAS~13224-3809 is dominated
below $\sim 2$ keV by a soft excess and from 2 to 10 keV by a hard
($\Gamma \sim 1.3$) power law (Otani \markcite{26} 1995).  In
contrast, a very steep spectrum with $\Gamma_{(2-10keV)} \sim 2.6$ was
found from NLS1 object RE~1034+39 (Pounds, Done \& Osborne
\markcite{39} 1995).

We report the results from December 1993 {\it ROSAT} and {\it ASCA}
observations of Mrk~766. Timing analyses of two {\it ROSAT} archival
observations are also presented. In section 2 the data reduction is
discussed briefly.  In section 3 timing analyses using normalized
variability amplitudes and hardness ratios are presented.  In section
4 the spectral analysis of the {\it ASCA} data is described.  The
results are discussed in terms of standard models in Section 5 and
compared with reported results from other NLS1s.  A summary and
conclusions are given in Section 6.

\section{{\it ASCA} and {\it ROSAT} Observations of Mrk~766}

We observed Mrk~766 with {\it ASCA} and the {\it ROSAT} PSPC during
December 1993.  It had been previously observed with the {\it ROSAT}
PSPC several times. Two longer observations were made 1991 June 15 and
1992 December 21 and these data were extracted from the {\it ROSAT}
archive. The observation log is given in Table 1.

The data were reduced using Xselect. To ensure all soft photons were
collected, {\it ROSAT} extraction regions of $3 ^\prime$ for the
on-axis 1993 observation and $4^\prime$ for the off-axis 1991 and 1992
observations were used. Extraction of background subtracted light
curves from the events files was done using IDL software.  The light
curves from the off-axis 1991 and 1992 observations were corrected for
vignetting. In the 1991 observation, Mrk~766 was periodically occulted
by the detector rib so time periods in which the flux dropped to zero
because of occultation were excluded (e.g. Brandt et al.
\markcite{2} 1993).  A region of the same radius located diametrically
across the detector and subjected to the same good time interval
selection provided an approximately correctly normalized background.

{\it ASCA} data were reduced using standard selection and cleaning
criteria.  Background for the GIS was obtained from a source free
region in the GIS field of view approximately the same distance from
the optical axis as the source.  Background SIS spectra were obtained
from blank sky fields, while background count rates for the light
curves were determined from the edges of the SIS chips. The time
dependent gain shift in the SIS was accounted for by filling and using
PI columns.  Spectra were extracted using Xselect, and background
subtracted light curves were obtained from the cleaned events files
using IDL.

Figure~\ref{fig1}a shows the {\it ROSAT} PSPC and {\it ASCA} SIS0
light curves from the most recent, quasi-simultaneous 1993
observations.  Error bars represent $1\sigma$ statistical error. The
{\it ROSAT} observation started $\sim 20,000$ seconds before the {\it
ASCA} observation but was unfortunately cut short as the satellite
went into safe hold mode.  The {\it ROSAT} PSPC light curves from the
1991 and 1992 observations are shown in Figure~\ref{fig1}b.  Mrk~766
was brightest in the {\it ROSAT} band in 1992 at $<5 \rm\,
cnts\,s^{-1}$.  This is comparable to the flux level observed during
the {\it ROSAT} All Sky Survey observation (Molendi, Maccacaro \&
Schaeidt \markcite{20} 1993).  In order to make a direct comparison
with the other data, only the first $\sim 80$ks of the 1992
observation was considered.  The remainder consists of data from only
two orbits, is separated from the main part of the observation by
nearly one day and the full light curve is shown in Netzer, George \&
Turner \markcite{24} 1994.

\section{Timing  Analysis}

\subsection{Fastest Observed Variability}

The {\it ASCA} light curves were examined in order to find instances
of rapid variability that could be clearly identified in all four
detectors.  A dip in flux occurred at $\sim 50,000\rm \, s$ from the
start of the observation; the light curve from the SIS0 detector is
shown embedded in Figure~\ref{fig1}a.  At the end of the dip, the flux
increased by nearly a factor of two in $\sim 1000$ s. This timescale
is the same order as those observed in {\it ASCA} data from other
Seyfert 1 nuclei, including NGC 4051 ($\Delta t \sim 200 \rm \,s$;
Mihara et al. \markcite{18} 1994).

Assuming that the X-ray emission originates from a single region, the
minimum flux doubling time scale $\Delta t$ gives an upper limit on
the source size as $R \sim c \Delta t$ of $3 \times 10^{13}\, \rm cm$,
where $c$ is the speed of light.  This estimate breaks down if the
X-rays are emitted from many small regions.

Because of the telescope wobble, variability on timescales less than
$400\rm \, s$ generally cannot be detected in {\it ROSAT} data.
Significant variability was observed consistently between adjacent
orbits ($\Delta t \sim 6000$ s).  The fastest variability of the three
observations was found when it was brightest during the 1992
observation.  A 30\% increase in 2400 s occurred 3400 s from the
beginning of the observation (Figure~\ref{fig1}b).

\subsection{Variance analysis}

The normalized variability amplitude (NVA), defined to be the standard
deviation divided by the mean intensity, provides a simple way to
quantify the variability in different energy bands (e.g. Edelson
\markcite{5} 1992).  If the width of the energy band is chosen so that the
number of photons in each light curve is the same, the NVA collapses
to the square root of the variance.

The {\it ASCA} light curves from each detector were accumulated with
100 s binning in the 4--10 keV band where the power law component
should dominate the spectrum.  The light curves were not background
subtracted.  However, the dilution of the variability by the constant
background was small since the background rate in the highest energy
band where contribution is largest was only e.g. $\sim 8$\% of the
SIS0 count rate.  The data below 4 keV were divided into energy bands
with bounds chosen so that the light curves had the same mean square
measurement error $\sigma^2_{err}$ as in the 4--10 keV band.  Note
that each resulting band was wider than the energy resolution at that
band.  The true variance of the data given by
$\sigma^2_{int}=\sigma^2_{obs}-\sigma^2_{err}$ is plotted as a
function of energy in Figure~\ref{fig2}.  The $1
\sigma$ uncertainties in the variance are less than 10\% of the values
(Equation 2 of Done et al.  \markcite{4} 1992).  These variance plots
demonstrate that the variability amplitude over the whole observation
is smallest at hard energies, peaks near 1 keV, and decreases towards
lower energies in the SISs.

For the {\it ROSAT} light curves, three standard energy bands were
considered: the `{\it a} band' (channels 11--41, energy 0.1--0.4 keV),
the `{\it c} band' (channels 52--90, energy 0.5--0.9 keV) and the
`{\it d} band' (channels 91--201, energy 0.9--2 keV). These bands are
independent and roughly correspond to regions where different spectral
features will dominate: the soft excess in the {\it a} band, the warm
absorber in the {\it c} band, and the power law in the {\it d} band.
The binning of the background subtracted light curves was chosen to be
400~s to account for the telescope wobble.  

Because of the rib occultations in the 1991 data, it was necessary to
include data with net exposure per bin of 200--400~s, even though the
shorter exposure bins add noise to the light curve. The light curves
in these three bands are shown on the left in Figures~\ref{fig3}a, b
and c.  Mrk~766 is bright enough and the bin size is long enough that
the signal to noise is better than 6 in all bins. Variability was
detected in each energy band and the \ch values for a constant
hypothesis model fit and the computed NVAs are listed in Table 2. The
NVAs show that in all observations the source is significantly less
variable in the lowest energy band compared with the higher energy
bands.

\subsection{Flux Ratios}

The hardness ratio (4.0--10.5 keV/1.0--1.35 keV) and softness ratio
(0.4--0.7 keV/1.0--1.35 keV) light curves from the {\it ASCA} SIS0
detector are shown in Figures~\ref{fig4}a and 4b, respectively.
Following Ptak et al.  \markcite{27} (1994) the values were computed
using variable bin sizes corresponding to good time intervals longer
than 300 seconds.  A large decrease in hardness corresponding to a
softening of the spectrum was observed in the hardness ratio light
curve at about 20,000~s from the beginning of the observation
(Figure~\ref{fig4}a).  The light curves show that the spectral change
is due to a large increase in 1.0--1.35 keV flux while the 4--10 keV
flux remained nearly constant.  Significant variability in the
softness ratio was also observed at the same time, such that the
spectrum below 1 keV hardens when the flux increases
(Figure~\ref{fig4}b).  The hardening of the spectrum may be more
gradual and it is not clearly completed until after $\sim 25,000$~s.
Variability was observed in both energy bands but the amplitude is
larger in the 1.0--1.35 keV band.  The spectral variability is
confined to the region around 20,000~s elapsed time.  Other instances
of large amplitude flux variability occurred (e.g. at $\sim 35,000$ s)
with no corresponding spectral change.  Thus, the spectral variability
is not strictly correlated with the flux.

{\it ROSAT} softness ratio light curves (0.1--0.4 keV/0.9--2.0 keV and
0.5--0.9 keV/0.9--2.0 keV) are shown on the right sides of
Figures~\ref{fig3}a, b and c.  Spectral variability is most pronounced
between the hardest and softest bands, with the spectrum generally
hardening as the flux increased.  This result is also true from
observation to observation; the spectrum was hardest when the source
was brightest in 1992, and softest when it was dimmest in 1991.  On
long time scales the softest band varies proportionately less than the
harder bands.  On short time scales the spectral variability is not
correlated with the flux.  At high flux, when the source was most
rapidly variable, correlated variability occurred with no change in the
hardness ratio (1992: at the beginning of the observation).  Similar
correlated variability was observed during the {\it ROSAT} All Sky
Survey when the source was also quite bright (Molendi, Maccacaro and
Schaeidt \markcite{20} 1993).  In contrast, at low flux in the 1991
and 1992 observations, orbit to orbit deviations in the hardest band
occurred which were not followed in the softest band (1991: at $\sim
23000$ and $\sim 58000$~s; 1992: at $\sim 66000$~s).  These short time
scale hard band excursions result in dips in the softness ratio.

\section{Spectral Analysis}

Based on the softness and hardness ratios, the {\it ASCA} data were
divided in two as shown in Figures~\ref{fig4}a and 4b, and spectra were
accumulated avoiding the transition region between $\sim 15,000$ and
$\sim 25,0000$~s from the beginning of the observation.  These spectra
are referred to as the low state and high state spectra and represent
exposures of $\sim 8,000$ and $20,000$~s respectively.
Figure~\ref{fig5} shows the pha ratios from the low state divided by
the high state of the summed SIS0 and SIS1 spectra in the top panel
and of the summed GIS2 and GIS3 spectra in the bottom panel.  These
confirm the results of the variance analysis which were that the
spectrum is most variable at around 1 keV, the variability amplitude
decreases to lower and higher energies, and the flux is essentially
constant at about 10 keV.

\subsection{The Hard X-ray Spectrum}

An estimation of the power law index, assuming that any warm absorber
does not have a large column density, was obtained by fitting the four
spectra above 2 keV with a power law plus iron $K\alpha$ line model.
The low and high state spectral indices were $1.57\pm 0.07$ and
$2.00^{+0.03}_{-0.04}$ respectively, and the results from fitting each
detector separately were consistent.  Throughout this paper, quoted
uncertainties are determined assuming 90\% confidence and 1 parameter
of interest ($\Delta \chi^2=2.71$).

\subsection{The Soft X-ray Spectrum}

The shape of the soft X-ray spectrum can be seen by fixing the power
law indices to the values found above 2 keV, including the Galactic
absorption ($1.77 \times 10^{20} \rm \, cm^{-2}$; Elvis, Wilkes \&
Lockman
\markcite{6} 1989), and plotting the ratio of the data to the model.
These plots are shown in Figure~\ref{fig6}a and 6b for the SIS0 and
GIS2 low state and high state spectra respectively.  In the low state
there is excess emission below 0.8 keV, suggesting the presence of a
soft excess component (e.g.  Mihara et al. \markcite{18} 1994).  In
the high state there is a deficit between 0.8 and 1.3 keV, suggesting
the presence of a partially ionized absorber (e.g. Fabian et al.
\markcite{8} 1994).  It is clear that an absorbed power law plus
narrow iron line cannot model the spectra.  Spectral fitting results
of the SIS0:GIS2 and SIS1:GIS3 pairs are listed in Table 3 for low and
high states separately.  The iron line is discussed separately in
Section 4.5 and fit results are given in Table 4.

The soft excess component and warm absorber can be simply
parameterized using a black body model and an absorption edge,
respectively.  The results from adding each of these components
separately are given in the second and third panels of Table 3.
Addition of either model component significantly improved the fits of
both the low and high flux spectra.  However, the addition of the
black body improved the fit more than the edge in the low flux case,
while the reverse was true for the high flux case, as expected from
the residuals shown in Figure~\ref{fig6}.

The fourth panel of Table 3 lists results from fitting with a model
including both a blackbody and an edge.  These fit results confirm the
importance of the black body component in the low state spectra and
the edge in the high state spectra. The black body temperatures and
edge energies of the low and high state spectra are consistent. The
edge energy near $0.74\rm \, keV$ is consistent with an origin of
transmission through a partially ionized absorber dominated by
\ion{O}{7}.  The power law slope of the low flux spectra remains
significantly flatter than that of the high flux spectra ($\Delta
\Gamma \sim 0.35$).

Next, to better simulate the warm absorber, a two edge plus power law
model was tried and the results are listed in the fifth panel of Table
3.  The power law indices are steep in both the low and high state,
and the implied change in index is much smaller ($\Delta\chi^2
\sim 0.1$).  This model has the same number of degrees of freedom as
the power law plus black body and edge model, but the fit is much
poorer for the low flux spectra ($\Delta \chi^2$ of 24 and 50 for the
SIS0:GIS2 and SIS1:GIS3 pairs, respectively).  In contrast, the fits
of the high flux spectra are improved somewhat by the additional edge
($\Delta \chi^2$ of 9 and 11).  Addition of a black body component,
while not necessary for the high flux spectra, greatly improves the
fit of the low flux spectra ($\Delta \chi^2$ of 28 and 52), and
results in a decrease in the low flux photon index and an increase in
the implied index change ($\Delta\Gamma \sim 0.35$).  The temperature
of the black body component in the low and high states is consistent
at $kT \sim 120\rm eV$, although the limits on the temperature in the
high state cannot be determined well for the SIS1:GIS3 pair.  The edge
energies near 0.74 and $0.87\rm \, keV$ found in the high state are
approximately consistent with absorption by \ion{O}{7} and \ion{O}{8}.
In the low state, the \ion{O}{7} edge is clearly detected but the
second edge is not necessary.  This suggests that the ionization of
the gas in the high state is higher than in the low state.

We also modeled the warm absorber using a table model (e.g.  Yaqoob,
Warwick \& Pounds \markcite{37} 1989).  The model used here assumes
that the power law is the sole source of ionizing photons. The density
of the gas was assumed to be $10^{9.5} \rm cm^{-3}$.  An analytic
approximation based on a large number of CLOUDY runs was used to model
the temperature as a function of ionization parameter.  It was found
to range between $\sim 5 \times 10^4$K and $\sim 10 \times 10^4$K. The
advantages of using the warm absorber table model compared with the
two edge model are that there are two fewer parameters, and the
parameters ($log(U)$ and $log(N_w)$) can be directly interpreted.  The
disadvantage is that we must assume a particular model for the
ionizing spectrum and gas.  Further, in this model, only absorption is
considered, and the emission lines expected if the warm absorber has a
large covering factor are ignored (e.g. Netzer \markcite{23} 1993).
However, this model is sufficient for a general discussion given the
statistical quality of these data (but see Section 4.3.1).

The four groups of spectra were first fit with a power law plus warm
absorber model (Table 3).  As found using the two edge model, the fits
of the high flux spectra are acceptable, but the fits of the low flux
spectra are unacceptable.  Addition of a black body component
substantially improves the fit of the low flux spectra, as shown in
the eighth panel of Table 3, but it is not necessary to model the high
flux spectra.  The ionized column density is consistent between the
low and high flux states at $log(N_w) \sim 21.8$, while the ionization
state $\log(U)$ is lower for the low flux spectra ($-0.78$) compared
to the high flux spectra ($-0.4$).  In this model, ionization states
$log(U)$ below and above $\sim -0.4$ are dominated by \ion{O}{7} and
\ion{O}{8} absorption, respectively. In contrast with the two edge
description of the warm absorber, the use of the warm absorber table
model resulted in a higher black body temperature in the high flux
state; however, the black body was barely detected in the high flux
spectra ($\Delta \chi^2$ of 4 and 5).  Thus the temperature of the
soft component is model dependent, but the necessity of this component
to model the low flux spectra is not. The table model description of
the warm absorber results in a slightly steeper index compared with
the two edge description.  This is because the table model properly
treats the absorption by gas with cosmic abundances (i.e. not only
oxygen) producing curvature in the model between 1.5 and 2.5 keV.
Thus the value of the indices is slightly model dependent, but the
change in index between the low and high states is not.

In summary, these results show that to model the high and low flux
spectra consistently, both a soft excess component and a warm absorber
are required.  When both of these components are included in the
spectrum, the low flux spectral index is significantly flatter than
the high flux index ($\Delta\Gamma \sim 0.35$).

\subsection{Other Models}

\subsubsection{Other Soft Excess Models}

The soft excess component in the low flux spectra was modeled
adequately with a black body (with a single edge, $\chi^2 = 230$/266
d.o.f. for the SIS0:GIS2 pair).  For comparison, other soft excess
models including Raymond--Smith (cosmic abundances), bremsstrahlung,
disk black body, and power law were tried.  The power law plus soft
component model alone did not fit the low flux spectra well.  The disk
black body model fit the best, with \ch of 247/268 d.o.f.  Including
an edge generally improved the fits, but the Raymond--Smith and the
power law models could not describe the low flux spectra well (\ch of
277 and 259/266 d.o.f. respectively), while the bremsstrahlung and the
disk black body models could (\ch of 234 and 231/266 d.o.f.
respectively).  A slightly higher temperature was found using these
models ($kT=200$ eV and $kT=145$ eV) compared with the black body
model ($kT=117$ eV) but nearly consistent edge energy, edge depth and
photon index were found.  The indices obtained were flat ($\Gamma =
1.57$ in both cases).

The warm absorber table model used thus far models only the absorption
edges of various ionized species.  Since the line emission expected
from a physical warm absorber might appear as an excess emission
component, we also tested models calculated using XSTAR which include
emission lines from the warm absorber (Kallman \& Krolik
\markcite{16} 1993).  Fitting the SIS0:GIS2 pair with a power law
plus emission only from a physical warm absorber in the line of sight
resulted in a better fit with \ch of 292/277 d.o.f.  Including an
edge, to simulate the absorption by a warm absorber, resulted in a \ch
of 261/268 d.o.f.  However, a significantly better fit was found when
a blackbody was also included (\ch = 228/264 d.o.f.) and the resulting
power law index was again flat ($\Gamma = 1.55$).  Similarly, emission
from reflection by a warm absorber could not alone describe the
spectrum (alone: $\chi^2=299/268$ d.o.f.; with an edge:
$\chi^2=268/266$ d.o.f.; with an edge and a black body:
$\chi^2=227/264$ d.o.f.).  In the final case, the index was flat
($\Gamma = 1.55$).

Soft excess emission can also be produced by reflection from an
ionized disk (Ross \& Fabian \markcite{28} 1993; Zycki et al.
\markcite{38} 1994).  We fit the low flux spectrum with an ionized
disk table model computed according to Zycki et al. \markcite{38}
1994.  For power law plus disk emission only the \ch was 262/269
d.o.f., but the ratio of the reflected flux to direct emission was
5.8.  Since the ratio should be near 1 for the isotropic static case.
we considered this result to be unphysical.  Addition of an edge gave
\ch of 248/267 d.o.f., but again the ratio was too large at 5.5.
Addition of a black body gave $\chi^2=228/265$ d.o.f., with the ratio
reduced to a physical value of $R=1.23$ and a flat index $\Gamma =
1.61$.

These results show that soft excess models without line emission fit
the spectra well, but we cannot distinguish among them, possibly
because of the poor statistics due to the relatively short exposure in
the low state and the decrease in sensitivity toward low energies of
the {\it ASCA} SIS.  Further, the flat spectral index obtained in the
low state is robust against changes in the soft excess model.

\subsubsection{Partial Covering Models}

Another possible origin of soft emission is leakage through a
partially covering absorber (e.g. NGC 4151: Weaver et al.
\markcite{35} 1994).  The spectral variability would then result from
a change in the covering fraction.  However, the partial covering
model does not fit the low flux SIS0 and GIS2 spectra well
($\chi^2=298$ for 268 d.o.f.), the resulting power law is forced to be
steep ($\Gamma=2.42$), and both low and high energy residuals are
seen.  These can be modeled by reflection, in which case the fit is
good ($\chi^2=225$/263 d.o.f.), but the power law is very steep
($\Gamma=3.0$) and the ratio of the reflected emission to primary
emission is required to be 20.  This model is unphysical so we
conclude that partial covering cannot adequately model the soft excess
component. Finally, a decrease in the fraction of the source covered can
only produce a steepening of the spectrum with an increase in flux and
cannot explain the hardening of the spectrum below 1 keV indicated by
the softness ratio (Figure~\ref{fig4}b).

A scattering and dual absorber model was used to describe the complex
X-ray spectrum of NGC 4151 (Weaver et al.  \markcite{35} 1994).  For
Mrk~766 this model does not give a good fit ($\chi^2=244$/266 d.o.f.)
and the photon index is steep ($\Gamma=2.84$). Low energy residuals
suggest an unmodeled absorption edge.  When an edge is included the
fit is good (\ch=222/264 d.o.f.), but the power law index is very
steep ($\Gamma=2.94$).  Again, this steep index seems unphysical, so
the dual absorber model is rejected.

\section{Reflection}

The spectral index change is robust against changes in models of the
soft excess component and warm absorber because the low flux spectrum
is flat at high energies.  The reflection spectrum is also flat (e.g.
George \& Fabian  \markcite{9} 1991), and could produce a hard tail if
the reflection component normalization is high compared with the power
law normalization.  This could be observed if the response of the
reflection component lags variability of the incident X-rays and would
be expected if the light crossing timescale of the reflection region
is long compared with the source variability timescale, or if the
reflection region is located far from the X-ray source (e.g. in the
molecular torus; Ghisellini, Haardt
\& Matt \markcite{12} 1994; Krolik, Madau \& Zycki
\markcite{17} 1994; Leighly et al. \markcite{42} 1996).

We define the reflection ratio to be 1 under the conditions that
nonvarying primary power law emission from an isotropic point source
illuminates an infinite optically thick disk.  In this case, the
reflection spectral component is not important in the spectrum below 5
keV.  Thus, we can estimate the contribution of the reflection by
fitting the spectra below 5 keV and comparing with the fit results
over the full range.  A power law plus two edge fit of the SIS0:GIS2
spectra results in a steep photon index (low state: $\Gamma=1.95$ and
\ch = 214.4/230 d.o.f., high state:  $\Gamma=2.02$ and \ch=524/519
d.o.f.).  Addition of a black body component to the model improved the
fit of the low flux spectra and flattened the photon index
($\Gamma=1.67$ and $\chi^2=197/228$ d.o.f.), but had no effect on the
high flux spectral fit.  If the table model is used to model the warm
absorber, similar results are obtained although both indices are found
to be slightly steeper.  Thus, fitting below 5 keV shows that the
index variability is still required and also that the measured low
energy index is consistent with the data above $\sim 5 \rm keV$.

A reflection ratio much larger than 1 results in a flat spectrum
towards high energies.  Fitting the spectra with a power law plus 2
edges and reflection allowing the ratio to be free produces a good fit
with a steep index which is consistent between the low and high flux
spectra (Table 3).  However, to explain the low flux spectrum, the
reflection ratio must be 5--8, while the high flux spectra require a
reflection ratio of only 0.5.  Most importantly, the reflection
component normalization is required to be significantly {\it higher}
in the low flux state compared with the high flux state, so a
reflection lag cannot explain the spectral variability.  Further, such
a large reflection ratio should be accompanied by a large equivalent
width narrow iron line in the low state spectrum and such a line is
not observed (Section 4.5).

The timing analysis results also rule out a lag in neutral reflection as
the origin of the spectral variability.  Since the neutral reflection
spectrum flux decreases towards low energies, only spectral softening
with an increase in power law flux is predicted, whereas a hardening
of the spectrum below 1 keV was observed. If the surface of the
reflector is ionized, the opacity is reduced at low energies and a
soft X-ray reflection component plus emission lines should be observed
(Ross \& Fabian \markcite{28} 1993; Zycki et al. \markcite{38} 1994),
which results in hardening of the low flux spectrum with an increase
in power law flux.  It was shown in Section 4.3.1 that the ionized
disk model does not produce a good fit alone, mostly because the soft
excess does not show evidence for line emission.

\subsection{The Iron Line}

The presence of the iron $K\alpha$ line in Seyfert 1 nuclei spectra
is well established (e.g. Nandra \& Pounds \markcite{22} 1994).  Mrk
766 was not observed using Ginga, and a iron $K\alpha$ line had not
yet been observed in its spectrum.  To look for the iron line, we fit
the spectra from all four detectors simultaneously above 2 keV.
Because of the continuum spectral variability the low and high state
spectra were fit separately.  A power law model resulted in
$\chi^2$/d.o.f. of 206/253 and 806/759 for the low and high states
respectively.  Addition of a narrow ($\sigma=0$ keV) redshifted line
with the energy fixed at 6.4 keV gave the results presented in Table
4.  There is no strong evidence for a narrow line with rest energy 6.4
keV in the low flux spectra ($\Delta\chi^2 \sim 2$). In the high flux
spectra $\Delta\chi^2$ is 13 corresponding to an F statistic value of
12.4, indicating the presence of a line with confidence greater than
99.9\%. Thus the presence of an iron emission line is confirmed in
this source.  The ratio of data to power law model for the summed SIS0
and SIS1 spectra shows the narrow line and ionized iron edge
(Figure~\ref{fig7}).  The non-detection in the low flux spectra can be
explained by the poor statistics resulting from lower flux and shorter
exposure.  If the high flux data are divided into several spectra
characterized by shorter exposures, the presence of a line in the
spectra from all detectors separately cannot be confirmed.  Next, the
line energy was allowed to be free.  The high state line energy is
consistent with 6.4 keV and a lower energy line is marginally detected
in the low state spectra.  The line energy in both states is
consistent with an origin in primarily neutral material, and emission
from highly ionized material is excluded. The line equivalent width is
$\sim 100 eV$ consistent with that expected from emission from an
accretion disk (e.g. George
\& Fabian  \markcite{9} 1991).  Broad iron lines have been discovered in the
{\it ASCA} spectra of several AGN (e.g.  Mushotzky et al.
\markcite{21} 1995).  A broad line is preferred in the low flux
spectra, although the addition of another parameter reduces the \ch by
only 5, so again the detection is marginal. The line is narrow in the
high flux spectra, constrained with $\sigma < 0.2\,\rm keV$.  The
shape of the high energy continuum changes the measured line
parameters.  When reflection with ratio fixed to 1 is included,
similar results were obtained as before, but the measured equivalent
widths were smaller (Table~4).

These results also support the hypothesis that a lag in the reflection
component cannot be the origin of the spectral variability.  A lag
implies that the bulk of the reflected emission should come from a
substantial distance from the source, and so the line would be
expected to be narrow.  However, the large reflection ratio required
to fit the low flux spectra predicts a very large equivalent width ($>
\sim 500\rm eV$) narrow line, which would be easily detected if
present.

\subsection{Quantifying the Spectral Variability}

The results of the previous sections indicate that the power law with
black body and warm absorber model provides the best fit to both the
low and high flux spectra.  A change in spectral index seems to be
indicated.  However, since the model is complex, spectra from the two
states must be fit simultaneously to determine which parameters
necessarily change.  Neutral absorption, originating in our Galaxy and
the host galaxy, was assumed not to change, and the iron line was
modeled as narrow with fixed energy.  Combined fits were done using both
descriptions of the warm absorber.

When two edges were used to model the warm absorber, the edge energies
were equated in the high and low state models.  This was done because
the low state spectra could not constrain the higher edge energy and
since the edges are identified as \ion{O}{7} and
\ion{O}{8} edges, no change is expected in the energies.  This model fit the
low and high spectra well (\ch of 860.1/859 d.o.f. and 789.8/844
d.o.f. for the SIS0:GIS2 and SIS1:GIS3 spectra respectively).  The
temperatures of the black body were consistent between the low and
high states so these were equated resulting in a negligible increase
in $\chi^2$.  No other parameters could be equated without resulting
in a large change in $\chi^2$.  The results are listed in the top
panel of Table~5, and they indicate that a significant change in index
and black body normalization occurred.  The edge energies are roughly
consistent with absorption by \ion{O}{7} and \ion{O}{8}.  The optical
depths of these edges are consistent between the low and high state,
so no change in the ionization of the warm absorber can be determined
from these fits.  However, the best fit value of $\tau_{OVII}$ is
larger in the low state than in the high state, and the reverse is
approximately true for $\tau_{OVIII}$.  This suggests that an increase
in the ionization occurred, but the statistics are too poor to require
this conclusion.

The results of the combined fits using the warm absorber table model
are listed in the second panel of Table 5.  The ionized column
densities were consistent so they were equated in the spectral
fitting.  As noted previously, when the warm absorber was described
using the table model, the black body temperature was found to be
higher in the high state than in the low state (Table 3).  When the
temperatures are equated in the combined fits, the increase in
$\chi^2$ is 2.7 and 5.7 for the SIS0:GIS2 and SIS1:GIS3 respectively,
significant with 90\% and 97.5\% confidence. However, this effect is
clearly model-dependent and may be due to the shape of the warm
absorber model and possibly the absence of emission lines in the
model, and thus the implied black body temperature change is unlikely
to be physical.  Three parameters changed between the low and high
states: the power law index, the black body normalization, and the
ionization parameter.  As these parameters are coupled in the spectral
fitting, to evaluate the significance of the change, the \ch contours
were plotted for each pair of parameters (Figure~\ref{fig8}a, b, and
c).  These show that the results are consistent between the SIS0:GIS2
and SIS1:GIS3 spectral pairs, and that the index change, the
ionization state change and the black body normalization change are
significant with $>99$\%, 90\% and 68\% confidence respectively.  We
note that the change in the ionization state is a model dependent
result, as we cannot demonstrate a change in the optical depths of the
two oxygen edges.  Figure~\ref{fig9}a and b show the best fitting
models, spectra, and ratios between spectra and model for the low and
high state SIS0:GIS2 spectra.  Figure~\ref{fig10} shows the best fit
models for the low and high state SIS0 spectra.  Note that the pivot
point for the power law change is $\sim 9$ keV.

\subsection{{\it ROSAT} Spectral Fitting}

The {\it ASCA} data show that the soft X-ray spectrum is complex,
comprised of a power law with variable index, a warm absorber with
variable ionization state and a black body with marginally variable
normalization and model dependent temperature.  Seven parameters are
required to describe the spectrum in the {\it ROSAT} band.  Because of
the poor spectral resolution, the PSPC spectra have 5 independent
channels.  Thus, detailed fitting of the PSPC spectra is of limited
value, as multiple models are acceptable.  Qualitatively, the spectra
from the 1992 and 1993 observations are very soft and cannot be
adequately modeled using a single power law plus absorption.  A soft
component like a black body gives a good fit, with $kT
\sim$70--90 eV, and a power law with index $\sim $1.9--2.0.  An edge
can also model the spectrum, but the spectral index is steeper at
$\sim 2.5$.

\section{Discussion}

\subsection{The Hard Spectral Variability}

The most significant result of this study was the observation of
dramatic photon index variability from $\sim 1.6$ to $\sim 2.0$, over
several thousand seconds and confined to a single event.  We discuss this
result in light of current models of the X-ray power law emission in AGN.

\subsubsection{General Considerations}

The high energy power law observed from AGN can be successfully and
plausibly explained by inverse Comptonization by high energy electrons
of soft UV photons likely originating in the accretion disk. The rapid
X-ray variability of some AGN implies a high radiation density in the
nucleus which results in production of electron-positron pairs. The
importance of pair production is determined by the compactness
parameter, $$l= L\sigma_T/R m_e c^3,\eqno(1)$$ where $L$ is the
luminosity, $R$ is the source size, $\sigma_T$ is the Thompson
scattering cross section, $m_e$ is the mass of the electron, and $c$
is the speed of light.  If the compactness is high, the optical depth
to pair production will exceed unity and pairs will be produced which
can substantially modify the emerging spectrum.  Generally speaking,
these models can be differentiated by whether the high energy
electrons are thermal or accelerated by non-thermal processes, since
pair production limits the highest energy attainable in the thermal
plasma.  In rapidly variable AGN, however, both thermal and
non-thermal processes may be present (Ghisellini, Haardt \& Fabian
\markcite{10} 1993).

The results presented here suggest that Mrk~766 is compact enough that
pairs should be produced in the nucleus.  The X-ray flux was observed
to change by a factor of two in $\sim 1000$ seconds. This corresponds
to a source size upper limit of $R < c\Delta t \sim 3 \times 10^{13}
\rm \, cm$.  The 2--10 keV luminosity is $1.3 \times 10^{43}\, \rm ergs \,
s^{-1}$ in the high state, implying an X-ray compactness parameter of
$l_x \sim 12$.  The hard compactness parameter, proportional to the
total luminosity in the hard component, could be substantially larger.
If the compactness parameters are larger than 10, pair production
should be important if there are an adequate number of $\gamma$-ray
photons present.  In non-thermal models, the $\gamma$-rays are
produced through upscattering of soft photons by extremely
relativistic electrons.  In thermal models, the origin is primarily the
high energy tail of the thermal spectrum and thus the number of
$\gamma$-rays depends on the temperature of the plasma. OSSE
observations of a few AGN find that the temperature is large enough
that electron positron pairs should be produced (e.g. see Figure 1 of
Fabian  \markcite{7} 1994); however, there have been no high energy
spectra obtained from Mrk~766.

\subsubsection{Simple Thermal Comptonization Models}

If the source does not contain many pairs, and if the power law
results from unsaturated Comptonization of soft photons, the spectral
parameters are very simply related to one another (e.g. Rybicki \&
Lightman \markcite{29} 1979).  The slope of the power law depends
inversely on the Compton $y$ parameter which is proportional to a
power of the temperature, so an increase in temperature implys a
decrease in index.  However, if the soft photon input is constant, the
pivot point energy of the photon index change should be the energy of
the soft input photons.  In contrast, we observe the pivot point to be
much higher, at $\sim 9$ keV.

If the thermal plasma is pair dominated, Ghisellini \& Haardt (1994)
\markcite{11} show that there is a one--to--one mapping of the
observables ($kT$ and $\alpha$, where $\alpha=\Gamma-1$ is the energy
index) to the plasma parameters ($l_H$ and $l_H/l_S$, where $l_H$ and
$l_S$ are the hard and soft compactnesses, characteristic of the
relativistic electrons and the soft (UV) seed photons, respectively).
Our observed increase in photon index by $\Delta \alpha \sim 0.4$
implies a decrease in $l_H/l_S$ by a factor of 10 (Figure 2 of
Ghisellini \& Haardt \markcite{11} 1994).  We observed a 2--10 keV
flux increase by a factor of 1.3, but since the power law pivot point
is $\sim 9\rm keV$, integration to high energies may show that the low
flux state luminosity is actually larger than the high flux state
luminosity.  OSSE observations of several Seyfert 1 galaxies have
found that the power spectrum is cut off above several hundred keV
(e.g. Fabian \markcite{7} 1994).  Integration of the power law from
2~keV to the generous upper limit of 500~keV shows that the low state
$l_H$ is at most a factor of 2 larger than the high state $l_H$,
predicting an increase in index by only $\sim 0.1$.  Further, the time
scale of the spectral variability, less than 10,000 seconds, precludes
a large increase in $l_S$, since this short time scale would be the
order of the orbital period at the innermost stable orbit for a $< 5
\times 10^{7} M_\odot$ black hole.  A change in the accretion rate
should be characterized by the viscous or radial drift time scale,
estimated by Molendi \& Maccacaro (1994) \markcite{19} to be 2.6 days.
Further, the {\it ROSAT} spectral variability can be most naturally
explained by a constant (on time scales of one day) soft component
dominating the softest X-ray band (see Section 5.2).  Finally, as
Ghisellini \& Haardt (1994) \markcite{11} note, if reprocessing in the
disk is important, $l_H/l_S$ would be expected to remain approximately
constant, and little intrinsic index variability should be observed.
However, this model is very simple, and predictions may change
substantially if a realistic geometry or self-consistent treatment of
the two phases is considered.

\subsubsection{Nonthermal Comptonization Models}

Static nonthermal models of X-ray emission have been investigated by
several authors (e.g. Svensson \markcite{31} 1994 and references
therein) and 2--10 keV spectral index variability has been studied by
Yaqoob \markcite{36} (1992). Most simply and generally, soft photons
with dimensionless frequency $x_S$ and compactness $l_S$ are scattered
by relativistic electrons with Lorentz factor $\gamma_0$ and
compactness $l_H$.  First order scattering produces a flat photon
spectrum with $\Gamma \sim 1.5$ extending to
$x_{max,1}=max[4/3\gamma_0^2 x_s,\gamma_0]$ (Svensson \markcite{30}
1987). Pairs are produced if the photon spectrum extends to
sufficiently high energies and if the optical depth to pair production
is greater than unity.  Soft photons reprocessed by pairs have a
steeper spectrum with $\Gamma=1.75$ breaking sharply at $x_B=2
\gamma_0^4 (2/3 x_s)^3$. If the energy of pair reprocessed photons is
relatively low, they could be observed as an X-ray soft excess.  If
the energy of reprocessed photons is high enough, additional pair
generations will be produced resulting in a pair cascade.  In that
case, the photon spectrum is steep with $\Gamma$ approaching 2 (e.g.
Svensson \markcite{30} 1987).

A nonthermal model can naturally explain the observed change in index,
the disappearance of the soft excess component and the confinement of
the spectral variability to a single event. The photon index
variability could result from a transition from a first order pair
spectrum to a cascade caused by a sudden increase in the Lorentz
factor of the relativistic electrons.  Thus, the low flux spectrum is
comprised of the inverse Compton cooling spectrum, characterized by
the hard X-ray power law with photon index near $1.5$, and the first
order optically thin pair reprocessed spectrum, observed as the soft
excess component.  The high flux spectrum is comprised of a pair
cascade spectrum, characterized by the hard X-ray power law with
photon index near 2.  The soft excess component disappears in the high
flux spectrum as the maximum energy of the pair reprocessed spectrum
increases far beyond the observed X-ray band, resulting in the
cascade.  The change in flux in the X-ray band depends on the change
of hard compactness $l_H$ which can be expected to accompany the
change in Lorentz factor, and flux variability uncorrelated with
spectral variability would occur through variation in $l_H$ alone.

A possible difficulty with non-thermal models which produce flat
spectra is that, in general, they overpredict the gamma ray
background.  However, recent work shows that if the non-thermal plasma
is in a corona above a disk, $\gamma$-ray photons are more efficiently
depleted and, depending on the compactness, a strong spectral cut off
below 200 keV is predicted (Tritz \markcite{32} 1990; Tsuruta \&
Kellen \markcite{40} 1995).  

\subsection{The Soft Spectral Variability}

The {\it ROSAT} spectrum was observed to become harder as the flux
increased, consistent with the lower amplitude variability observed at
softest energies. A most natural explanation for this behavior is
variability between the relative normalizations of the power law and
soft excess component.  This would be observed if the flux of the soft
component were nearly constant on the time scale of an observation.
In terms of current physical models it could imply that the soft
component is dominated by primary emission from an accretion disk and
reprocessed hard emission is relatively less important.  The fraction
of black body flux in the softest band can be estimated by comparing
the relative variability of the hardest band where the power law
dominates with the relative variability of the softest band where the
soft component dominates.  This scenario predicts that if a static
soft excess component comprises a large fraction of the flux in the
softest band, any flux change must be accompanied by a hardness ratio
change, while if the black body comprises only a small fraction of the
flux, variability in hard and soft bands should be correlated.

Overall, the {\it ROSAT} PSPC data are qualitatively consistent with
this scenario. In the 1992 observation, the spectrum was harder at
high flux and correlated variability was observed, while at low flux
uncorrelated variability was found.  In the 1991 observation, when the
flux was lower and the spectrum generally softer, only uncorrelated
variability was observed.  This scenario can also explain the lack of
spectral variability seen in the {\it ROSAT} All Sky Survey data
(Molendi, Maccacaro and Schaeidt \markcite{20} 1993) since at that
time the source was bright and power law emission may have dominated
the soft component emission.  Qualitatively and on short time scales
there are difficulties with this scenario.  In 1992, the hardest and
softest bands decrease by 50\% and 25\% respectively overall, implying
the soft component must contribute the same percentages of the soft
band flux in the high and low states respectively.  However, at high
flux there is a dip in flux by 30\% just after the start of the
observation, and no change in the softness ratio was observed.
Similarly, at low flux, there is an increase by 30\% in the hard band
approximately $6\times 10^4\,\rm s$ after the start of the observation,
but no change in the soft band emission was observed.  In combination,
these two results cannot be explained if the black body flux is constant
and no other parameters change.  Reprocessing, neglected so far, may
be able to explain large amplitude correlated variability at high
flux.

Other models cannot explain the observed spectral variability.
Netzer, Turner \& George (1994) \markcite{24} showed that variability
of the warm absorber in response to ionizing flux changes could not
explain the spectral variability found during the 1992 observation.
In general, if the soft component were dominated by reprocessing, the
soft X-ray variability would be expected to track the hard component
variability.  However, substantial spectral variability of the
incident continuum could result in some spectral variability of the
reprocessed component.  Changes in the accretion rate (Molendi \&
Maccacaro \markcite{19} 1994) cannot explain the orbit-to-orbit
spectral variability as the predicted time scales are much longer.

\subsection{The Change in the Warm Absorber}

In the {\it ASCA} spectra, we found that there was no evidence that
the ionized absorption column density changed between the high state
and the low state; however, we found that the ionization parameter
changed with 90\% confidence.  This result is model dependent, as we
could not demonstrate that the optical depth of the oxygen edges
changed significantly.

The best fit ionization parameter $\log(U)$ changed from $\sim -0.85$
in the low state to $\sim -0.42$ in the high state, implying an
increase in flux of ionizing photons by a factor of 2.7.  It is
interesting that this is quite close to the implied change in flux by
a factor of 3.1 of the intrinsic power law at 0.7 keV. The observed
photon index variability implies a larger change in the flux of
ionizing photons, assuming the power law extends to low energies.  The
spectral changes occurred over a time scale of several thousand
seconds.  The recombination time scale for \ion{O}{8} is about $2
\times 10^{11}T_e^{0.5} n^{-1}$ seconds (e.g.  Turner et al.
\markcite{33} 1993) or about 5.5 hours using the parameters assumed in
this model (or longer if the gas is rarer).  Thus the gas may not be
in photoionization equilibrium in the high flux state, and the larger
population of \ion{O}{8} implied by the increase in the ionization
parameter may result from ions which are directly stripped of an
electron by the increased number of photons with energy near 0.7 keV.
Further observations are necessary to determine the response of the
ionized material to a decrease in flux since a change in ionization
would be observed only if recombination had occurred.

A change in the ionization correlated with an increase in flux has not
been previously reported from {\it ASCA} data.  An increase in column
density and no change in ionization accompanied an increase in flux in
MCG--6-30-15 (Fabian et al. \markcite{8} 1994).  Explaining the
spectral variability during an increase in flux by a change in warm
absorber properties in NGC~3227 required a decrease in ionization and
an increase in column (Ptak et al. \markcite{27} 1994). In another
observation of MCG--6-30-15, an increase in the optical depth of the
\ion{O}{8} edge during a flux decrease was interpreted as evidence for
recombination of \ion{O}{9} (Otani \markcite{26} 1995).

\subsection{Hard X-ray Emission from Narrow Line Seyfert 1s}
\vskip 1pc

The soft X-ray properties of narrow line Seyfert 1s are well studied
(Boller, Brandt \& Fink \markcite{1} 1996) but few hard X-ray
observations have been reported. The {\it ASCA} observation of Mrk~766
represents one of the first observations of the hard emission from
these objects.

We found a hard power law with variable photon index in Mrk~766, and
the spectral variability which was not strictly flux correlated.
Similar photon index variability which was flux correlated has been
discovered from NGC~4051 (Guainazzi et al. \markcite{15} 1996), a
Seyfert 1 galaxy which shares many properties with NLS1s.  In
contrast, a very steep spectrum with photon index $\sim 2.6$, a
dominate soft excess and no variability was observed from narrow-line
Seyfert 1 RE~1034+39 (Pounds, Done \& Osborne \markcite{39} 1995).
Because these spectral and variability properties are similar to those
characteristic of black hole candidates in the high state (e.g. Nowak
\markcite{41} 1991) it was postulated that RE~1034+39 represents a
Seyfert 1 galaxy in the high state (Pounds, Done \& Osborne
\markcite{39} 1995).

The marked differences between the hard X-ray properties of Mrk~766
and RE~1034+39 are interesting.  Black hole candidates in the low
state are characterized by a flat hard X-ray power law and more rapid,
larger amplitude hard X-ray variability (e.g. Nowak \markcite {41}
1995).  These properties more closely resemble the observational
results from Mrk~766 and NGC~4051 than do the properties of black hole
candidates in the high state.  Further observations of NLS1s may find
that the spectral and variability properties fall into two classes:
those with steep hard X-ray spectra, dominant soft X-ray emission and
lower amplitude short term hard X-ray variability, and those with flat
hard X-ray spectra, less soft X-ray emission and rapid hard X-ray
variability.  There is perhaps already some evidence for such a
division. While many NLS1s are very bright soft X-ray objects commonly
found in soft X-ray samples, relatively few have hard X-ray detections
by HEAO-1 A2.

Further support for this scenario may come if repeated X-ray
observations discover that some objects have made the transition
between two states.  This could have been what had occurred in objects
observed to have varied by factors of 10 or more between two {\it
ROSAT} observations (e.g. Zwicky 159.034, Brandt, Pounds \& Fink
\markcite{3} 1995; WPVS007, Grupe et al. \markcite{14} 1995).
Well-studied broad-line Seyfert 1 galaxies are not observed to undergo
this kind of transition.  The fact that many well-studied AGN are hard
X-ray selected while NLS1s are clearly soft X-ray selected objects
further supports this hypothesis.

In black hole candidates, it is widely believed that the high state
is characterized by a relatively larger accretion rate compared with
the low state.  Thus the behavior of NLS1s may result from a
relatively larger accretion rate.  If the processes fueling AGNs are
common for Seyferts, a relatively larger accretion rate would be
observed in systems with relatively small black hole masses.

\section{Conclusions}

We report analysis of {\it ASCA} and {\it ROSAT} observations of the
narrow-line Seyfert 1 galaxy Mrk~766.  In the {\it ASCA} observation
rapid variability with doubling time scale of order $\sim 1000$
seconds was observed, and dramatic spectral variability over as time
period of less than $\sim 10,000\,\rm s$ was discovered. Confined to a
single event, during a 2--10 keV flux increase the spectrum above and
below $\sim 1$ keV softened and hardened respectively. The low and
high flux spectra could be described with a model consisting of a
power law, iron line, warm absorber and soft excess modeled as a black
body.  The spectral variability was a result of a highly significant
increase in the intrinsic power law index from $\sim 1.6$ to $\sim
2.0$ with the pivot point at $\sim 9$ keV, a model dependent increase
in the ionization of the warm absorber, and a marginal decrease in the
soft excess component.  A $100 \rm \, eV$ equivalent width narrow iron
line was detected in the high flux spectrum but not in the low flux
spectrum, most likely because of poor statistics.  Variability on time
scales as short as $\sim 2400$ seconds was found in the {\it ROSAT}
data.  Because the variability in the softest {\it ROSAT} band, below
0.4 keV, had relatively lower amplitude than the harder bands,
spectral hardening during flux increases was detected on time scales
as short as the orbital period of $ \sim 6000 \,\rm s$.

The spectral index change, the disappearance of the soft component in
the {\it ASCA} band and the confinement of the spectral variability to
a single event could be naturally explained in terms of non-thermal
Comptonization models.  We postulate that the index change occurred
through a transition from a first order pair reprocessed spectrum to a
pair cascade spectrum brought about by a sudden increase in the
Lorentz factor of the injected relativistic electrons.  The first
order pair reprocessed spectrum observed in the low state as a soft
excess disappeared in the high state cascade spectrum. Variations in
the hard compactness resulted in pure flux variability.  The measured
increase in the warm absorber ionization corresponds to the increase
in flux near the oxygen edges resulting from the power law index
change.  The spectral variability in the {\it ROSAT} data was most
naturally explained by a variable hard component and a nonvariable
soft component which dominated the softest band and may be primary
emission from an accretion disk perhaps implying that reprocessing is
relatively less important in this object.

The flat and variable hard power law index observed in Mrk~766 is
similar to that observed in NGC 4051 (Guainazzi et al. \markcite{15}
1996), a Seyfert 1 with many properties common to NLS1s, but contrasts
markedly with the very steep hard X-ray index $\Gamma \sim 2.6$ found
in NLS1 object RE~1034+39 (Pounds, Done \& Osborne \markcite{39} 1995).
Further hard X-ray observations of NLS1s using {\it ASCA} are
necessary to clearly understand the hard X-ray properties of these
sources.

\acknowledgements

This research has made use of data obtained through the High Energy
Astrophysics Science Archive Research Center Online Service, provided
by the NASA-Goddard Space Flight Center. The authors thank T. Kallman
and P. Zycki for the use of their spectral table models.  KML
acknowledges receipt of a National Research Council fellowship at
NASA/GSFC and a Japanese Science and Technology Agency fellowship at
RIKEN and useful conversations with M. Cappi.  KML acknowledges
support by a {\it ROSAT} AO4 guest observer grant and an {\it ASCA}
AO1 guest observer grant.

\clearpage

\begin{deluxetable}{lrrrrr}
%\tablewidth{33pc}
\tablewidth{0pc}
\tablenum{1}
\tablecaption{Observing Log}
\tablehead{
\colhead{Instrument}   & \colhead{Observation} & 
\colhead{Exposure}     & \colhead{Total Counts} &
\colhead{Background}   \\
\colhead{}             & \colhead{Date} &
\colhead{(s)}          & \colhead{} &
\colhead{Count Rate}  }

\startdata
{\it ROSAT} PSPC & 15/06/91 & 15179\tablenotemark{a} & 21804 & 0.081   \nl
{\it ROSAT} PSPC & 21/12/92 & 16303\tablenotemark{b} & 53865 & 0.056  \nl
{\it ROSAT} PSPC & 17/12/93 & 3146 & 7309 & 0.033  \nl
{\it ASCA} S0 & 18/12/93 & 32947 & 32823 & 0.030  \nl
\hphantom{{\it ASCA}} S1 & & 32735 & 26853 & 0.020 \nl
\hphantom{{\it ASCA}} G2 & & 35805 & 16991 & 0.033 \nl
\hphantom{{\it ASCA}} G3 & & 35808 & 19532 & 0.032 \nl
\tablecomments{Column 4 is the total (not background subtracted) counts in the
source extraction region.  
Column 5 is the background rate scaled to the source
extraction region.  }
\tablenotetext{a} {Exposure time after data selection to remove periods
where the source was occulted by a rib.}
\tablenotetext{b} {Exposure time in the first 85,000 seconds of
observation.  The total observation time was 19870 seconds.}
\enddata
\end{deluxetable}
\clearpage

\begin{deluxetable}{lrrr}
%\tablewidth{33pc}
\tablewidth{0pc}
\tablenum{2}
\tablecaption{PSPC Variability}
\tablehead{
\colhead{Band}   & \colhead{Mean} & 
\colhead{$\chi^2_\nu$}     & \colhead{NVA}}

\startdata

\multicolumn{4}{l}{1991 Data (44 points):} \nl
Total & 1.35 & 17.6 & 0.22 \nl
0.2--0.5 & 0.84 & 6.0 & 0.16 \nl
0.5--0.9 & 0.20 & 7.05 & 0.36 \nl
0.9--2.0 & 0.22 & 8.54 & 0.37 \nl
\tableline
\multicolumn{4}{l}{1992 Data (41 points):} \nl
Total & 3.06 & 77.5 & 0.26 \nl
0.2--0.5 & 1.63 & 22.7 & 0.19 \nl
0.5--0.9 & 0.58 & 24.9 & 0.34 \nl
0.9--2.0 & 0.67 & 35.1 & 0.38 \nl
\tableline
\multicolumn{4}{l}{Quasi-simultaneous} \nl
\multicolumn{4}{l}{Data (8 points):} \nl
Total & 2.24 & 29.9 & 0.19 \nl
0.2--0.5 & 1.33 & 8.2 & 0.13 \nl
0.5--0.9 & 0.39 & 13.9 & 0.31 \nl
0.9--2.0 & 0.39 & 13.6 & 0.31 \nl
\enddata
\end{deluxetable}
\clearpage

\begin{deluxetable}{lcccc}
\footnotesize
%\tablewidth{33pc}
\tablewidth{0pc}
\tablenum{3}
\tablecaption{{\it ASCA} Spectral Fitting Results}
\tablehead{
\colhead{Parameter}   & \multicolumn{2}{c}{Low State} &
\multicolumn{2}{c}{High State} \\
\colhead{} & \colhead{SIS0:GIS2} & 
\colhead{SIS1:GIS3} & \colhead{SIS0:GIS2} &
\colhead{SIS1:GIS3}}

\startdata
\multicolumn{5}{l}{Power law model:} \nl
$N_H (\times 10^{21}\rm cm^{-2})$ &
 $\rm Gal < 0.19$ & $\rm Gal<0.20$ & $\rm Gal<0.19$ & $\rm Gal<0.192$ \nl
Index & $1.65 \pm 0.04$ & $1.64^{+0.04}_{-0.05}$ & $1.92 \pm 0.02$ & 
$1.91^{+0.01}_{-0.02}$ \nl
\chs/d.o.f. & 422/270 & 341/254 &
 819/598 & 765/597 \nl
\tableline
\multicolumn{5}{l}{Power law plus black body:} \nl
$N_H (\times 10^{21}\rm cm^{-2})$ & $0.84^{+0.51}_{-0.48}$ & 
$0.75^{+0.53}_{-0.49}$
& $0.95^{+0.22}_{-0.20}$ & $1.02^{+0.22}_{-0.20}$ \nl
Index & $1.54^{+0.06}_{-0.07}$ & $1.54^{+0.08}_{-0.07}$ & $1.96 \pm 0.03$ &
$1.98^{+0.04}_{-0.03}$ \nl
SIS PL norm\tablenotemark{a}  & $3.0^{+0.3}_{-0.2}$ & $3.0^{+0.3}_{-0.2}$ &
$7.7 \pm 0.03$ & $7.9 \pm 0.03$ \nl
kT (eV)  & $88^{+8}_{-6}$ & $87^{+9}_{-7}$ & $77^{+5}_{-4}$ & $72 \pm 5$ \cr
SIS bb norm\tablenotemark{b} & $2.5^{+2.1}_{-1.2}$ & $2.3^{+2.1}_{-1.2}$ &
$4.2^{+1.5}_{-1.3}$ & $5.4^{+2.0}_{-1.7}$ \nl
\chs/d.o.f. & 245/268 & 208/254 & 681/596 & 647/595 \nl
\tableline
\multicolumn{5}{l}{Power law plus edge:} \nl
$N_H (\times 10^{21}\rm cm^{-2})$ & $\rm Gal < 0.18$ & $\rm Gal<0.20$ & $\rm Gal<0.26$ &
$\rm Gal< 0.23$ \nl
Index & $1.87 \pm 0.04$ & $1.83^{+0.06}_{-0.05}$ & $2.01 \pm 0.02$ &
$2.01^{+0.03}_{-0.02}$ \nl
SIS PL norm\tablenotemark{a}
 & $4.38 \pm 0.18$ & $4.23^{+0.24}_{-0.23}$ & $8.06^{+0.17}_{-0.15}$
& $8.06^{+0.23}_{-0.17}$ \nl
Edge Energy (keV) & $0.81 \pm 0.02$ & $0.82^{+0.02}_{-0.03}$ &
$0.78^{+0.01}_{-0.03}$ & $0.75 \pm 0.02$ \nl
$\tau$ & $1.01 \pm 0.15$ & $0.89^{+0.21}_{-0.19}$ & $0.49^{+0.07}_{-0.06}$ &
$0.53 \pm 0.70$ \nl
\chs/d.o.f. & 344/268 & 279/254 & 651/596 & 608/595 \nl
\tableline
\multicolumn{5}{l}{Power law, black body and edge:} \nl
$N_H (\times 10^{21}\rm cm^{-2})$ & $\rm Gal<0.28<0.91$ & $\rm Gal<0.42<0.88$ &
$0.43^{+0.18}_{-0.20}$ & $\rm Gal<0.35<0.63$ \nl
Index & $1.55^{+0.09}_{-0.05}$ & $1.57 \pm 0.07$ & $1.99^{+0.02}_{-0.04}$ &
$2.00^{+0.04}_{-0.03}$ \nl
SIS PL norm\tablenotemark{a} & $3.1^{+0.3}_{-0.2}$ & $3.2 \pm 0.30$ &
 $7.9^{+0.2}_{-0.3}$ & $8.0^{+0.4}_{-0.3}$ \nl
Edge energy (keV) & $0.76^{+0.03}_{-0.04}$ & $0.74 \pm 0.02$ &
$0.74^{+0.03}_{-0.02}$ & $0.75^{+0.02}_{-0.03}$ \nl
$\tau$ & $0.60^{+0.28}_{-0.25}$ & $0.77^{+0.28}_{-0.30}$ & 
$0.43^{+0.09}_{-0.10}$ & $0.46^{+0.12}_{-0.11}$ \nl 
$kT$ (eV) & $117 \pm 18$ & $121^{+21}_{-16}$ & $101^{+19}_{-13}$
 & $91^{+48}_{-35}$ \nl
SIS bb norm\tablenotemark{b}
 & $0.95^{+1.19}_{-0.26}$ & $1.12^{+0.82}_{-0.44}$ & $0.81^{+0.56}_{-0.50}$
& $0.54^{+1.04}_{-0.52}$ \nl
\chs/d.o.f. & 230/266 & 191/253 & 640/594 & 605/593 \nl
\tableline
\multicolumn{5}{l}{Power law and 2 edges:} \nl
$N_H (\times 10^{21}\rm cm^{-2})$ & $\rm Gal<0.22$ & $\rm Gal<0.22$ & $1.1<2.8$ &
$2.5<3.4$ \nl
Index & $1.91^{+0.05}_{-0.04}$ & $1.90 \pm 0.05$ & $2.02^{+0.05}_{-0.01}$ &
$2.05 \pm 0.04$ \nl
SIS PL norm\tablenotemark{a} & $5.0 \pm 0.20$ & $4.9 \pm 0.30$ & 
$8.2^{+0.5}_{-0.2}$ & $8.5^{+0.5}_{-0.4}$ \nl
Edge energy (keV) & $0.78^{+0.01}_{-0.02}$ & $0.77 \pm 0.03$ &
$0.74^{+0.02}_{-0.04}$ & $0.73 \pm 0.02$ \nl
$\tau$ & $1.00^{+0.18}_{-0.15}$ & $1.01^{+0.23}_{-0.18}$ &
 $0.43^{+0.08}_{-0.19}$ & $0.51^{+0.09}_{-0.07}$ \nl
Edge energy (keV) & $1.19^{+0.05}_{-0.03}$ & $1.24^{+0.07}_{-0.14}$ &
$0.94^{+0.06}_{-0.24}$ & $0.98^{+0.05}_{-0.04}$ \nl
$\tau$ & $0.55 \pm 0.11$ & $0.49^{+0.15}_{-0.11}$ & $0.16^{+0.15}_{-0.06}$ &
$0.14^{+0.07}_{-0.06}$ \nl
\chs/d.o.f. & 254/266 & 241/252 & 631/594 & 594/593 \nl
\tableline
\tablebreak
\multicolumn{5}{l}{Power law, 2 edges and black body:} \nl
$N_H (\times 10^{21}\rm cm^{-2})$ & $Gal<0.32<0.81$ & $\rm Gal<0.70$ &
$\rm Gal<0.29<0.44$ & $\rm Gal<0.49<0.52$ \nl
Index & $1.64^{+0.10}_{-0.09}$ & $1.62 \pm 0.09$ & $2.01 \pm 0.04$ &
$2.05 \pm 0.04$ \nl
SIS PL norm\tablenotemark{a} & $3.5^{+0.5}_{-0.4}$ & $3.4^{+0.4}_{-0.3}$ &
$8.2 \pm 0.4$ & $8.5^{+0.5}_{-0.4}$ \nl
Edge energy (keV) & $0.76 \pm 0.03$ & $0.75^{+0.02}_{-0.03}$ &
$0.73^{+0.02}_{-0.04}$ & $0.73 \pm 0.02$ \nl
$\tau$ & $0.67^{+0.26}_{-0.27}$ & $0.84^{+0.37}_{-0.30}$ & 
$0.40^{+0.15}_{-0.16}$ & $0.51^{+0.08}_{-0.07}$ \nl
$kT$ (eV) & $115 \pm 21$ & $134^{+31}_{-33}$ & $126^{+34}_{-29}$ & 
84\tablenotemark{c} \nl
SIS bb norm\tablenotemark{b} & $0.88^{+1.00}_{-0.37}$ &
$0.69^{+1.04}_{-0.13}$ & $0.29^{+0.31}_{-0.26}$ & $0<0.08$ \nl
Edge energy (keV) & $1.20^{+0.07}_{-0.17}$ & 1.104\tablenotemark{c}
 & $0.87^{+0.12}_{-0.06}$ & $0.98^{+0.05}_{-0.04}$ \nl
$\tau$ & $0.18^{+0.16}_{-0.14}$ & $0<0.16<0.27$ & $0.18^{+0.14}_{-0.09}$ &
$0.14 \pm 0.07$ \nl
\chs/d.o.f.  & 226/264 & 189/250 & 629/592 & 594/591 \nl
\tableline
\multicolumn{5}{l}{Warm Absorber:} \nl
$N_H (\times 10^{21}\rm cm^{-2})$ & $\rm Gal<0.18$ & $\rm Gal < 0.18$ &
$\rm Gal<0.23<0.30$ & $0.26^{+0.09}_{-0.08}$ \nl
Index & $2.05^{+0.04}_{-0.07}$ & $1.98 \pm 0.08$ & $2.12^{+0.05}_{-0.04}$ &
$2.13^{+0.05}_{-0.04}$ \nl
SIS pl norm\tablenotemark{a} & $6.4^{+0.5}_{-0.7}$ & $5.7^{+0.5}_{-0.3}$ &
9$.7 \pm 0.5$ & $9.9 \pm 0.6$ \nl
$log(U)$ & $-0.17^{+0.08}_{-0.19}$ & $-0.28^{+0.06}_{-0.09}$ &
$-0.38^{+0.05}_{-0.06}$ & $-0.43^{+0.08}_{-0.06}$ \nl
$log(N_w)$\tablenotemark{c} & $22.27^{+0.09}_{-0.17}$ & $22.13 \pm 0.09$ &
$21.71 \pm 0.07$ & $21.68^{+0.08}_{-0.06}$ \nl
\chs/d.o.f. & 294/268 & 269/254 & 634/596 & 601/595 \nl
\tableline
\multicolumn{5}{l}{Warm Absorber plus black body:} \nl
$N_H (\times 10^{21}\rm cm^{-2})$ & Gal<0.48<0.94 & $\rm Gal<0.36<0.79$ &
$\rm Gal<0.29<0.42$ & $\rm Gal<0.23<0.40$ \nl
Index & $1.70^{+0.13}_{-0.11}$ & $1.70^{+0.14}_{-0.10}$ &
 $2.12^{+0.05}_{-0.07}$ & $2.10^{+0.07}_{-0.06}$ \nl
SIS PL norm\tablenotemark{a} & $3.9^{+1.0}_{-0.6}$ & $4.0^{+1.1}_{-0.6}$ &
$9.8^{+0.9}_{-0.8}$ & $9.5^{+1.0}_{-0.6}$ \nl
$log(U)$ & $-0.78^{+0.21}_{-0.32}$ & $-0.79^{+0.25}_{-0.26}$ & 
$-0.41^{+0.08}_{-0.07}$ & $-0.42^{+0.06}_{-0.05}$ \nl
$log(N_w)^{d}$ & $21.78^{+0.36}_{-0.39}$ & $21.84^{+0.40}_{-0.36}$ & 
$21.82^{+0.13}_{-0.20}$ & $21.83 \pm 0.11$ \nl
$kT$ (eV) & $119^{+47}_{-18}$ & $131^{+55}_{-24}$ & $184^{+76}_{-103}$ &
$230^{+105}_{-57}$ \nl
SIS bb norm\tablenotemark{b}
 & $1.52^{+1.34}_{-0.66}$ & $1.37^{+1.70}_{-0.48}$ & $0.49^{+0.71}_{-0.46}$
& $0.49^{+0.65}_{-0.44}$ \nl
\chs/d.o.f. & 234/266 & 194/253 & 630/594 & 596/593 \nl
\tableline
\multicolumn{5}{l}{Power Law, two edges and Reflection:} \nl
$N_H (\times 10^{21}\rm cm^{-2})$ & $\rm Gal < 0.39$ & $\rm Gal < 0.50$ &
$\rm Gal<0.23<0.37$ & $\rm Gal<0.28<0.44$ \nl
Index & $2.05^{+0.12}_{-0.06}$ & $2.04^{+0.17}_{-0.06}$ & 
$2.06^{+0.08}_{-0.05}$ & $2.07^{+0.09}_{-0.06}$ \nl
SIS PL norm\tablenotemark{a} & $4.87^{+0.52}_{-0.22}$ & $4.75^{+0.83}_{-0.26}$ &
$8.40^{+0.61}_{-0.36}$ & $8.65^{+0.58}_{-0.51}$ \nl
SIS Refl norm\tablenotemark{a}
 & $2.73^{+1.83}_{-1.10}$ & $2.56^{+2.42}_{-0.94}$ & $0.21<1.24$ &
$0.26<1.35$ \nl
Refl. Ratio & $5.5^{+2.5}_{-2.1}$ & $8.3^{+6.1}_{-2.6}$ & $0.4<1.5$ &
$0.3<1.5$ \nl
Edge Energy (keV) & $1.18^{+0.05}_{-0.1}$ & $1.11^{+0.05}_{-0.06}$ &
$0.94^{+0.07}_{-0.11}$ & $0.99 \pm 0.04$  \nl
$\tau$ & $0.44^{+0.08}_{-0.12}$ & $0.38^{+0.13}_{-0.12}$ & 
$0.18^{+0.06}_{-0.08}$ & $0.15 \pm 0.07$ \nl
Edge Energy (keV) & $0.77 \pm 0.02$ & $0.75 \pm 0.02$ &
$0.73 \pm 0.02$ & $0.73^{+0.02}_{-0.01}$ \nl
$\tau$ & $0.98 \pm 0.17$ & $0.90^{+0.18}_{-0.17}$ & $0.44^{+0.10}_{-0.20}$ &
$0.53\pm 0.09$ \nl
\chs/d.o.f. & 230/263 & 201/250 & 631/593 & 594/590 \nl
\tablecomments{The value ``Gal'' for the absorption refers to the spectral
fit lower limit set to the Galactic value, $1.77 \times 10^{20} \rm \,
cm^{-2}$ (Elvis, Wilkes \& Lockman \markcite{6} 1989).}
\tablenotetext{a}{$\times
10^{-3} \rm photons\,keV^{-1}cm^{-2}s^{-1}$}
\tablenotetext{b}{$\times 10^{-4} L_{39}/{D_{10}^2}$, where 
$L_{39}$ is the source
luminosity in $10^{39}\rm erg\,s^{-1}$ and $D_{10}$ is the distance to
the source in $10^{10} \rm kpc$}
\tablenotetext{c}{Unconstrained}
\tablenotetext{d}{log of the ionized column in units of $\rm cm^{-2}$.}
\enddata

\end{deluxetable}
\clearpage

\begin{deluxetable}{lcc}
%\tablewidth{33pc}
\tablewidth{0pc}
\tablenum{4}
\tablecaption{Iron Line Fitting Results}
\tablehead{
\colhead{Parameter}   & \colhead{Low State} & 
\colhead{High State}}

\startdata
\multicolumn{3}{l}{Power law plus narrow line:} \nl
Index & $1.57 \pm 0.07$ & $2.00^{+0.03}_{-0.04}$ \nl
Line Flux\tablenotemark{a} & $1.8<3.0$ & $2.3^{+1.0}_{-1.2}$ \nl
Line Eq. Width (eV) & $100<170$ & $110^{+40}_{-55}$ \nl
\chs/d.o.f. & 204/252 & 793/758 \nl
\tableline
\multicolumn{3}{l}{Power law, narrow line and reflection:} \nl
Index & $1.66 \pm 0.07$ & $2.08^{+0.04}_{-0.03}$ \nl
Line Flux\tablenotemark{a}& $1.1<2.6$ & $1.8^{+0.9}_{-1.0}$ \nl
Line Eq. Width (eV) & $35<84$ & $47 \pm 25$ \nl
\chs/d.o.f. & 204/252 & 793/758 \nl
\tablenotetext{a}{$\times 10^{-5} \rm photons\,cm^{-2}s^{-1}$ in the
line.}
\enddata
\end{deluxetable}
\clearpage

\begin{deluxetable}{lcccc}
\footnotesize
%\tablewidth{33pc}
\tablewidth{0pc}
\tablenum{5}
\tablecaption{Combined Spectral Fits}
\tablehead{
\colhead{Parameter}   & \multicolumn{2}{c}{SIS0:GIS2}
   & \multicolumn{2}{c}{SIS1:GIS3} \\
\colhead{} & \colhead{Low State} &
\colhead{High State} & \colhead{Low State} &
\colhead{High State}}

\startdata
\multicolumn{5}{l}{Two Edge Model: } \nl
$N_H (\times 10^{21}\rm cm^{-2})$ & \multicolumn{2}{c}{$\rm Gal<0.27<0.45$} &
\multicolumn{2}{c}{$\rm Gal<0.25<0.37$} \nl
Index & $1.58 \pm 0.08$ & $2.02 \pm 0.04$ & $1.60^{+0.08}_{-0.07}$ &
$2.05 \pm 0.04$ \nl 
SIS PL norm\tablenotemark{a} & $3.2 \pm 0.30$ & $8.3^{+0.4}_{-0.3}$ 
& $3.2 \pm 0.3$ & $8.6 \pm 0.4$ \nl
kT (eV) & \multicolumn{2}{c}{$123^{+22}_{-17}$} &
 \multicolumn{2}{c}{$133^{+26}_{-21}$} \nl
SIS BB norm\tablenotemark{b} & 0$.89^{+0.34}_{-0.22}$ & $0.19<0.61$ & 
$0.84^{+0.24}_{-0.16}$ & $0<0.23$ \nl
Edge Energy (keV) & \multicolumn{2}{c}{$0.73^{+0.03}_{-0.01}$} &
\multicolumn{2}{c}{$0.74^{+0.01}_{-0.02}$} \nl
$\tau$ & $0.66^{+0.28}_{-0.25}$ & $0.47^{+0.08}_{-0.09}$ & 
$0.83^{+0.19}_{-0.17}$ & $0.52^{+0.07}_{-0.08}$ \nl
Edge Energy (keV) & \multicolumn{2}{c}{$0.97^{+0.05}_{-0.16}$} &
\multicolumn{2}{c}{ $0.99\pm 0.04$ } \nl
$\tau$ & $0.11<0.30$ & $0.15^{+0.06}_{-0.08}$ & $0.13<0.30$ & 
$0.14^{+0.07}_{-0.06}$\nl
\chs/d.o.f & \multicolumn{2}{c}{860.1/860} &
\multicolumn{2}{c}{790.7/845 } \nl 
\tableline 
\multicolumn{5}{l}{Warm Absorber Model: } \nl
$N_H (\times 10^{21}\rm cm^{-2})$ & \multicolumn{2}{c}{$0.33^{+0.16}_{-0.15}$} &
\multicolumn{2}{c}{$0.30^{+0.17}_{-0.10}$} \nl
Index & $1.66^{+0.07}_{-0.06}$ & $2.12^{+0.05}_{-0.04}$ &
$1.67^{+0.07}_{-0.06}$ & $2.14^{+0.05}_{-0.04}$ \nl
SIS PL norm\tablenotemark{a}
 & $3.7^{+0.2}_{-0.3}$ & $9.8 \pm 0.6$ & $3.6^{+0.3}_{-0.2}$ &
$10.0 \pm 0.6$ \nl 
kT (eV)  & \multicolumn{2}{c}{$117^{+13}_{-14}$} & 
\multicolumn{2}{c}{$121^{+12}_{-14}$} \nl
SIS BB norm\tablenotemark{b}
 & $1.34^{+0.43}_{-0.38}$ & $0.42<0.97$ & $1.20^{+0.49}_{-0.28}$ &
$0.02<0.52$ \nl
$log(N_w)$\tablenotemark{c} & \multicolumn{2}{c}{$21.68^{+0.04}_{-0.08}$} 
& \multicolumn{2}{c}{$21.69 \pm 0.08$}
\nl
$log(U)$ & $-0.85^{+0.16}_{-0.15}$ & $-0.42 \pm 0.08$  &
$-0.87^{+0.16}_{-0.15}$ & $-0.42^{+0.07}_{-0.09}$ \nl
\chs/d.o.f & \multicolumn{2}{c}{867.1/863} & \multicolumn{2}{c}{801.8/848} \nl
\tablecomments{The value ``Gal'' for the absorption refers to the spectral
fit lower limit set to the Galactic value, $1.77 \times 10^{20} \rm \,
cm^{-2}$ (Elvis, Wilkes \& Lockman \markcite{6} 1989).}
\tablenotetext{a} {$\times 10^{-3} \rm
photons\,keV^{-1}cm^{-2}s^{-1}$}
\tablenotetext{b} {$\times 10^{-4} L_{39}/{D_{10}^2}$, where $L_{39}$
 is the source
luminosity in $10^{39}\rm erg\,s^{-1}$ and $D_{10}$ is the distance to
the source in $10^{10} \rm kpc$.}
\tablenotetext{c}{log of the ionized column in units of $ \rm
cm^{-2}$}
\enddata
\end{deluxetable}

\newpage

\figcaption{{\it (a.)} The 1993 {\it ROSAT} PSPC (0.1--2 keV) and {\it ASCA}
SIS0 (0.4--10 keV) light curves with the same absolute time scale.
The  {\it ASCA} data were binned by good time intervals with
minimum interval of 300 seconds.  The binning for the {\it ROSAT} light
curves is 400 seconds, to account for the wobble of the telescope. The
SIS0 light curve from the period of most rapid variability is
embedded. {\it (b.)} The 1991 and 1992 {\it ROSAT} PSPC (0.1--2 keV) light
curves corrected for vignetting and on the same relative time
scale.\label{fig1}}

\figcaption{Variance versus energy of {\it ASCA} light curves for each
detector in energy bands with bounds chosen so that the mean square
measurement error $\sigma^2_{err}$ is the same in each.  These show
that the largest variability amplitude occurred at $\sim 1 \rm
keV$.\label{fig2}}

\figcaption{On the left, vignetting corrected {\it ROSAT} light curves
in three standard energy bands, and on the right, the softness ratio
light curves.  The time bin size was 200--400 seconds. {\it(a.)} 1991
observation; {\it (b.)} 1992 observation; {\it (c.)} 1993
observation.\label{fig3}}

\figcaption{{\it ASCA} flux ratios as a function of elapsed time are
shown in the upper panels while the light curves in each energy band
are shown in the lower panels.  In both cases the light curves were
computed using variable bin size corresponding to good time intervals
longer than 300 seconds (Ptak et al.  1994).  The data accumulated in
the high and low flux spectra are also marked.  {\it (a.)} The
hardness ratio (4.0--10.5 keV/1.0--1.35 keV); {\it (b.)} the softness
ratio (0.4--0.7 keV/1.0--1.35 keV).\label{fig4}}

\figcaption{Pha ratios of the low state versus high state spectra.
The upper and lower panels show the results from SIS0+SIS1 and
GIS2+GIS3 spectra, respectively.  These ratios confirm the variance
analysis results shown in Figure 2 that the spectrum is most variable
at around 1 keV, less variable towards higher and lower energies and
essentially constant at around 10 keV.  Further, the smoothly
increasing ratios from 2--10 keV suggest that a variable photon index
causes the spectral variability between these energies.\label{fig5}}

\figcaption{Residuals from a power law fit above 2 keV for the
SIS0:GIS2 pair accounting for Galactic absorption. {\it (a.)} In the
low state the index is flat ($\Gamma=1.56$) and excess emission is
found below 0.8 keV, suggesting a soft excess component.  {\it (b.)}
In the high state the index is steep ($\Gamma=2.01$) and a deficit is
found around 0.8 keV, suggesting an ionized absorber.\label{fig6}}

\figcaption{Ratio of data to model for a power law fit to the summed
SIS0 and SIS1 data above 2 keV showing the narrow iron line and
ionized iron absorption edge. \label{fig7}}

\figcaption{\ch contours for two degrees of freedom for the
combined power law, black body and warm absorber model fits. The
SIS0:GIS2 and SIS1:GIS3 contours are shown on the left and right
sides, respectively. {\it (a.)} Photon index versus black body
normalization, {\it (b.)} $log(U)$ versus photon index, and {\it (c.)}
$log(U)$ versus black body normalization.  The change in the photon
index is significant at $>99$\% confidence, the change in the
ionization parameter is significant at 90\% confidence and the change
in the black body normalization is significant at 68\%
confidence.\label{fig8}}

\figcaption{The best fit model (power law, black body and warm
absorber table), spectra and ratio of spectra to model for the SIS0:GIS2
pair, on the left and right sides respectively. {\it (a.)} the low
state, {\it (b.)} the high state.\label{fig9}}

\figcaption{The unfolded best fit model (power law, black body and
warm absorber table) showing additive model components for the SIS.
The low state and high state models are shown on the left and right
sides respectively.  Note the pivot point for the variable power law
index occurs near 9 keV, and that the large decrease in black body
normalization is marginally significant (see Figure 6).\label{fig10}}

\clearpage


\clearpage


\begin{references}

\reference{1} Boller, Th., Brandt, W. N., \& Fink, H. 1996, \aap, 305, 53

\reference{2} Brandt, W. N., Fabian, A. C., Nandra, K., \& Tsuruta, S.,
1993, \mnras, 255, 996

\reference{3} Brandt, W. N., Pounds, K. A., \& Fink, H. 1995, \mnras,
273, 47

\reference{4} Done, C., Madejski, G. M., Mushotzky, R. F., Turner,
 T. J., Koyama, K., \& Kunieda, H., 1992, \apj, 400, 138


\reference{5} Edelson, R. 1992, \apj, 401, 516

\reference{6} Elvis, M., Lockman, F. J., \& Wilkes, B. J. 1989, \aj, 97, 777

\reference{7} Fabian, A. C., 1994, \apjs, 92, 555

\reference{8} Fabian, A. C., et al. 1994, PASJ, 46, L59

\reference{9} George, I. M. \& Fabian, A. C. 1991, \mnras, 249, 352

\reference{10} Ghisellini, G., \& Haardt, F., \& Fabian, A. C., 1993,
\mnras, 263, L9

\reference{11} Ghisellini, G., \& Haardt, F., 1994, \apjl, 429, L53

\reference{12} Ghisellini, G., \& Haardt, F., \& Matt G., 1994,
\mnras, 267, 743

\reference{13} Goodrich, R. W., 1989, \apj, 342, 224

\reference{14} Grupe, D., Beuermann, K., Mannheim, K, Bade, N., 
Thomas, H.--C., de Martino, D., \& Schwope, A., 1995, \aap Letters, 300, 21

\reference{15} Guainazzi, M., et al., 1996, PASJ, in press

\reference{16} Kallman, T., \& Krolik, J. H. 1993, GSFC preprint

\reference{17} Krolik, J. H., Madau, P., Zycki, P. T., 1994, \apj,
420, 57

\reference{42} Leighly, K. M., Kunieda, H., Awaki, H., \& Tsuruta, S.,
1996, \apj, in press (May 20, Vol. 463)

\reference{18} Mihara, T., Matsuoka, M., Mushotzky, R. F., 
Kunieda, K., Otani, C., Miyamoto, S., \& Yamauchi, M., 1994, PASJ 46, L137

\reference{19} Molendi, S., \& Maccacaro, T., 1994, \aap, 291, 420

\reference{20} Molendi, S., Maccacaro, T., \& Schaeidt, S. 1993, \aap,
 271, 18

\reference{21} Mushotzky, R. F. et al. 1995, \mnras, 272, L9

\reference{22} Nandra, K. \& Pounds, K. A., 1994, \mnras, 268, 405

\reference{23} Netzer, H. 1993, \apj, 411, 594

\reference{24} Netzer, H., Turner, T. J. \& George, I. M. 1994, \apj,
 435, 106

\reference{41} Nowak, M. A., 1996, PASP, 107, 55

\reference{25} Osterbrock, D. E., \& Pogge, R. W., 1985, \apj, 297, 166

\reference{26} Otani, C. 1995, PhD thesis, Tokyo University

\reference{39} Pounds, K.A., Done, C., \& Osborne, J.P., 1995, \mnras,
277, 5P

\reference{27} Ptak, A., Yaqoob, T., Serlemitsos, P. J., Mushotzky,
 R., \& Otani, C., 1994, \apjl, 436, L31

\reference{28} Ross, R. R., \& Fabian, A. C., 1993, MNRAS, 261, 74

\reference{29} Rybicki, G. B. \& Lightman, A. P. 1979, 
``{\it Radiative Processes in
Astrophysics''}, (New York: Wiley)

\reference{30} Svensson, R. 1987, \mnras, 227, 403

\reference{31} Svensson, R. 1994, \apjs, 92, 585

\reference{32} Tritz, B, 1990, PhD thesis, Montana State University

\reference{40} Tsuruta, S., \& Kellen, M., 1995, \apjl, 453, 9

\reference{33} Turner, T. J., Nandra, K., George, I. M., Fabian,
 A. C., \& Pounds, K. A., 1993, \apj, 419, 127

\reference{34} Urry, C. M., et al. 1989, {\it Proc. 23rd ESLAB Symp.},
 eds. J. Hunt \& B. Battrick (Paris: ESA) p. 789

\reference{35} Weaver, K. A., Yaqoob, T., Holt, S. S., 
Mushotzky, R. F., Matsuoka,
M., \& Yamauchi, M., 1994, \apjl, 436, L27

\reference{36} Yaqoob, T. 1992, \mnras 258, 198

\reference{37} Yaqoob, T., Warwick, R. S., \& Pounds, K. A., 
1989, \mnras, 236, 153

\reference{38} Zycki, P. T., Krolik, J. H., Zdziarski, A. J.,
 \& Kallman, T. R., 1994, \apj, 437, 597

\end{references}
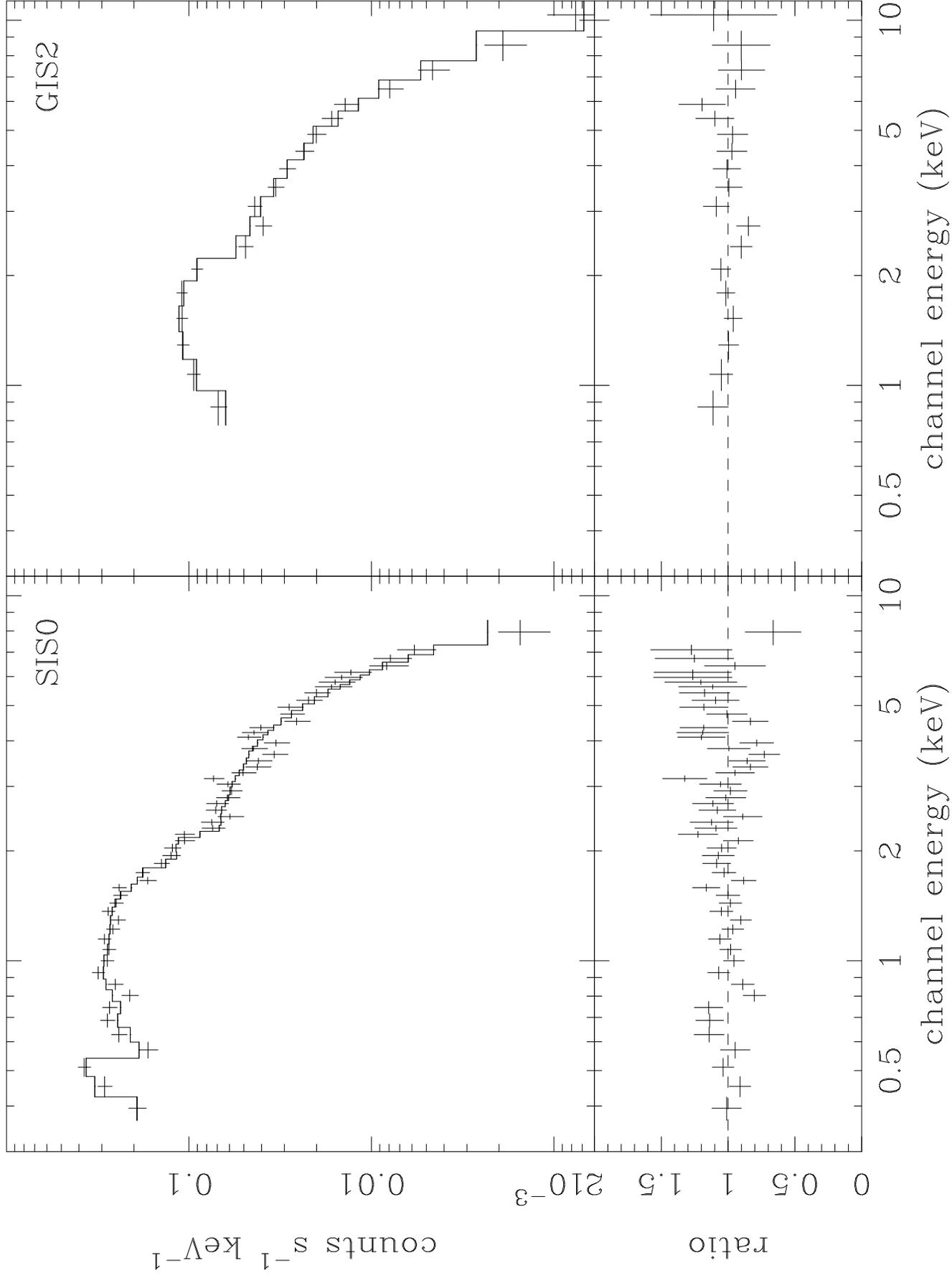
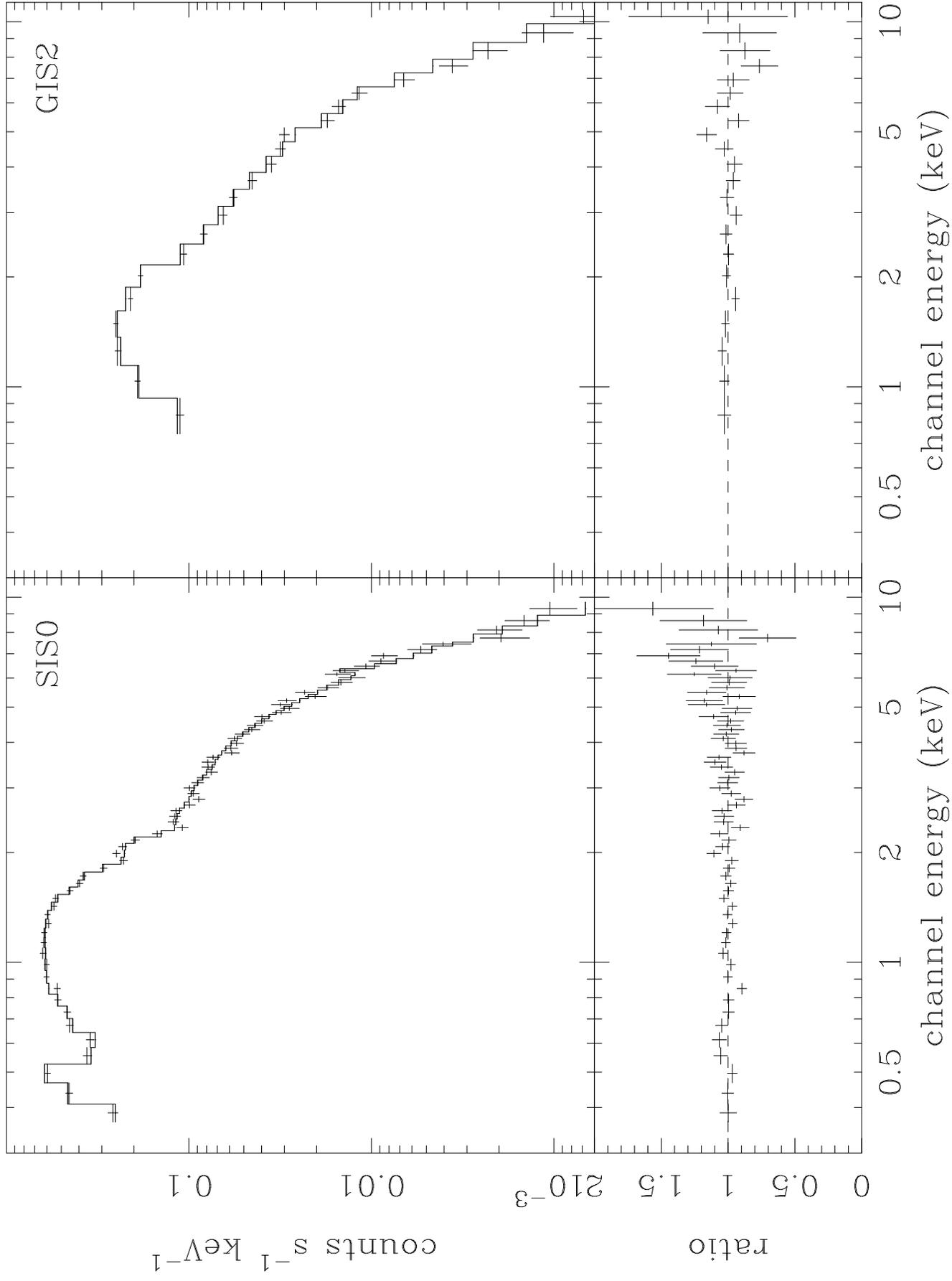
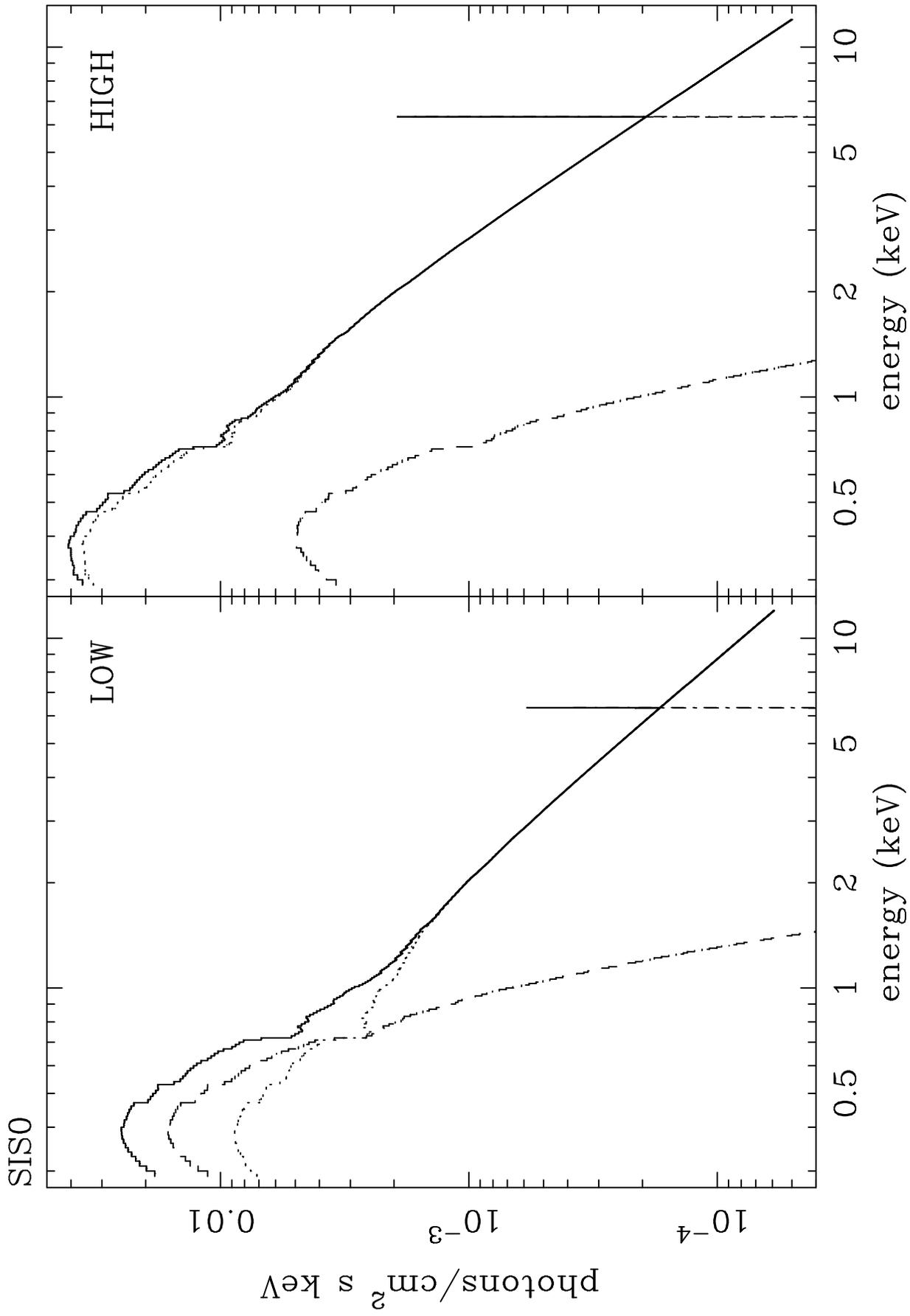
\end{document}